\documentclass[10pt,twocolumn,prl]{revtex4-1}
\usepackage[T1]{fontenc}
\usepackage[latin9]{inputenc}
\usepackage{xcolor}
\usepackage{babel}
\usepackage[normalem]{ulem}
\usepackage{amsmath}
\usepackage{amssymb}
\usepackage{graphicx}
\usepackage[colorlinks]{hyperref}


\begin{document}

\title{Confined Meson Excitations in Rydberg-Atom Arrays Coupled to a Cavity Field}
\author{Tharnier O. Puel$^{1,2}$ and Tommaso Macr\`{i}$^{3,2}$}
\affiliation{
$^{1}$Department of Physics and Astronomy, University of Iowa, Iowa City, Iowa 52242, USA}
\affiliation{$^{2}$Departamento de F\'{i}sica Te\'{o}rica e Experimental,  Universidade Federal do Rio Grande do Norte, Campus Universit\'{a}rio, Lagoa Nova, Natal-RN 59078-970, Brazil}
\affiliation{$^{3}$ITAMP, Harvard-Smithsonian Center for Astrophysics, Cambridge, Massachusetts 02138, USA
}

\begin{abstract}
Confinement is a pivotal phenomenon in numerous models of high-energy and statistical physics. In this study, we investigate the emergence of confined meson excitations within a one-dimensional system, comprising Rydberg-dressed atoms trapped and coupled to a cavity field. This system can be effectively represented by an Ising-Dicke Hamiltonian model. The observed ground-state phase diagram reveals a first-order transition from a ferromagnetic-subradiant phase to a paramagnetic-superradiant phase. Notably, a quench near the transition point within the ferromagnetic-subradiant phase induces meson oscillations in the spins and leads to the creation of squeezed-vacuum light states. We suggest a method for the photonic characterization of these confined excitations, utilizing homodyne detection and single-site imaging techniques to observe the localized particles. The methodologies and results detailed in this paper are feasible for implementation on existing cavity-QED platforms, employing Rydberg-atom arrays in deep optical lattices or optical tweezers.
\end{abstract}
\maketitle

{\it Introduction.}
The strong nuclear force bounds elementary particles with increasing strength with distance and appears between quarks to form, e.g., protons, neutrons, mesons, etc. Due to that, quarks are not found in isolation, a phenomenon called confinement. In condensed matter systems the confinement effect exists in low-dimensional-spin excitations~\cite{McCoy-1978,Fonseca-2003,Fonseca-2006,Kormos:2017ud,PhysRevLett.122.150601}, also referred to as meson excitations~\cite{Fonseca-2003}, and their presence are known to change qualitatively the spreading of correlations. Interestingly, such effect was experimentally observed in the CaCu$_2$O$_3$ crystals probed by neutron scattering measurements and mathematically modeled by a weakly-coupled spin ladder~\cite{lake-2010}.
Neutron scattering was later used to observe confinement in the quasi-one-dimensional ferromagnet CoNb$_2$O$_6$, and that could be well described by a transverse-field Ising model in the presence of a weak-longitudinal field~\cite{Coldea-2010}.
More recently, advances in quantum technology allowed the creation and manipulation of confined spin excitations in a trapped-ion quantum simulator~\cite{Tan:2021wf,RevModPhys.95.035002}.
Recent studies have extended this understanding to include scattering events and dynamical formation of novel hadronic states in quantum spin chains with long-range interactions, opening new avenues in the simulation of quantum chromodynamics using trapped-ion or Rydberg-atom setups 
\cite{PRXQuantum.3.040309, wang2023quantum}.

Our study introduces a novel approach to investigate confinement, employing a hybrid quantum device that utilizes a finite-range interacting spin chain coupled to a cavity field after a quantum quench. This method offers a fresh perspective on confinement dynamics, distinct from previous studies.
The experimental realization of the model involves neutral atoms, which are excited to Rydberg states via far-off-resonant processes \cite{PhysRevLett.104.195302,PhysRevA.89.011402,PhysRevA.107.062609,Jau-2016}.
This specific experimental setup has been proposed to explore and elucidate exotic many-body quantum phases
\cite{PhysRevA.87.061602,PhysRevLett.104.195302, PhysRevLett.128.113602,Macri-2014,Cinti-2014}.
Furthermore, this arrangement has shown promise for the generation of states with significant metrological applications \cite{PhysRevLett.116.053601,PhysRevA.94.010102,PhysRevLett.123.260505,PhysRevLett.124.063601,PhysRevLett.131.063401,PhysRevResearch.5.L012033,Eckner-2023}.
The confinement of these Rydberg atoms is achieved through their localization in optical lattices \cite{SciPostPhys.1.1.004} or within tweezer arrays with programmable geometries \cite{Bernien:2017aa,Labuan-2016}.
A pivotal aspect of this configuration is the interaction between the Rydberg atom arrays and a single mode of a cavity field which represents a significant advancement in understanding and controlling confined spin and photon excitations, a core novelty of our research.

{\it Model and Symmetries.}
The quantized transverse field Ising model (QTFIM)~\cite{PhysRevResearch.2.023131} combines two well-known models: the Ising and the Dicke model, $H_{\text{QTFIM}}=H_{\text{Ising}}+H_{\text{Dicke}}$.
The one-dimensional ($1$D) Ising model describes the exchange interaction between spins positioned along a chain
\begin{equation}
H_{\text{Ising}}=-J_{z}\sum_{n}\sigma_{n}^{z}\sigma_{n+1}^{z},
\label{eq:H Ising}
\end{equation}
where $J_z>0$ denotes the strength of the ferromagnetic interaction, and $\sigma_n^z$ represents the $z$-spin projection of the $n$-th spin.
The Dicke model describes the light-matter interaction, specifically a single photon mode interacting uniformly with the spin chain as
\begin{equation}
H_{\text{Dicke}}=\omega_{z}S_{z}+\frac{g}{\sqrt{N_{s}}}\left(a^{\dagger}+a\right)S_{x}+\omega_{a}a^{\dagger}a,
\label{eq:H Dicke}
\end{equation}
where $S_{\alpha}\equiv\sum_{i}\sigma_{i}^{\alpha}/2$ and $N_{s}$ represents the total
number of spins. 
The frequency splitting between atomic levels and the photon frequency are characterized by $\omega_z$ and $\omega_a$, respectively.
Note that, when combined, the $\omega_{z} S_z$ term acts as a longitudinal-magnetic field for the Ising model, whereas $g$ represents the strength of a transverse-magnetic field for the spins. The cavity field is characterized by the quantized term $\left(a^{\dagger}+a\right)$, which distinctively differentiates this from the well-known transverse-field Ising model (TFIM). 
A similar model for light-matter interaction was studied in Ref. \cite{Zhu_2019}, where, instead, the coupling acts as an effective longitudinal-magnetic field for the spins.

The TFIM exhibits a $Z_2$ symmetry, $\sigma_n^z \rightarrow -\sigma_n^z$, leaving the Hamiltonian invariant. This symmetry is explicitly broken in the presence of the longitudinal field $\omega_{z} S_z$. The Dicke model is invariant under the simultaneous transformation $a \rightarrow -a$ and $S_x \rightarrow - S_x$. This symmetry results in vanishing magnetization along the $x$-direction.

\begin{figure}
\hfill{}\includegraphics[width=0.90\columnwidth]{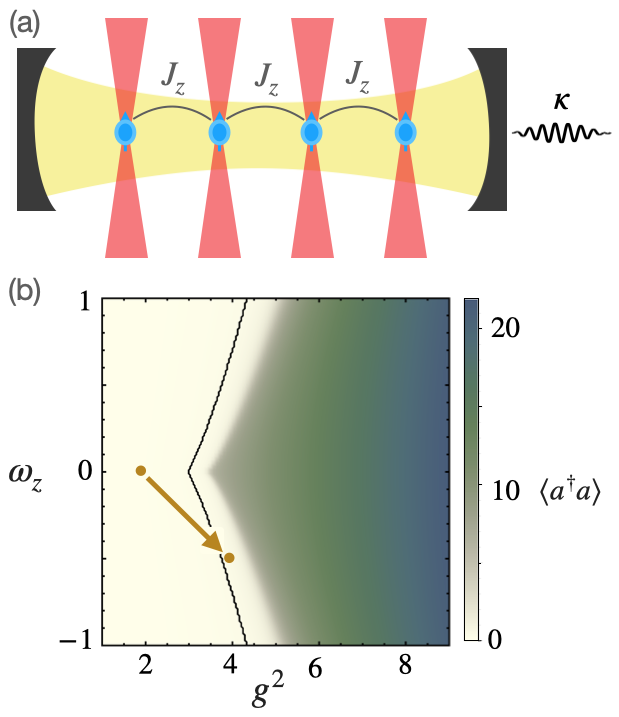}\hfill{}

\caption{(a) Illustration representing an array of atoms (blue) trapped by optical tweezers within an optical cavity with loss rate $\kappa$. (b) Phase diagram of the QTFIM for the spin-excitation frequency $\omega_{z}$
and spin-light interaction strength $g$. The colors represent the
photon number $\langle a^{\dagger}a\rangle$, computed by exact diagonalization
of $H_{\text{QTFMI}}$ for $N_{s}=10$ (truncating the photon basis to $N_{p}=50$). The photon number is the order parameter for the transition
between a ferromagnetic phase (with $\langle a^{\dagger}a\rangle=0$ and
$\left\langle \sigma_{r_{0}}^{z}\right\rangle =-1$) and a superradiant
phase (with $\langle a^{\dagger}a\rangle\protect\neq0$ and $\left\langle \sigma_{r_{0}}^{z}\right\rangle \protect\neq-1$).
The black line indicates the phase transition obtained from the mean-field
approach. 
The brown dots and arrow refer to the quench dynamics discussed in the main text.
}
\label{fig: phase diagram QTFIM}
\end{figure}

{\it Phase Diagram.} The TFIM undergoes a second-order phase transition, induced by the transverse field, where the $Z_2$ symmetry is spontaneously broken leading to a ferromagnetic phase \cite{PhysRevA.3.786}.
For vanishing $\omega_{z}$, the QTFIM resembles the TFIM, with the transverse field being quantized via the operators $(a^\dagger + a)$.
However, the QTFIM exhibits a first-order phase transition, whereby the increasing coupling strength $g$ causes the system to transition from a ferromagnetic-subradiant phase (characterized by the absence of photon creation) to a paramagnetic-superradiant phase (where photons are collectively created by all spins) \cite{PhysRevA.92.013624,PhysRevResearch.2.023131}. 
The complexity of the model increases substantially in the presence of an additional longitudinal field~\cite{Atas_2014,Atas_2017}, with a well-defined phase diagram only known for a random distribution of the field~\cite{wang-li-1994,CANKO200518,Suzuki-Inoue-Chakrabarti-2012}.
Contrastingly, we show that the same does not apply to the QTFIM. Indeed, the first-order phase transition remains robust in the presence of $\omega_{z}(\neq0)$.

From an exact diagonalization analysis of the $H_\text{QTFIM}$, we calculated the photon number $\langle a^\dagger a \rangle$ in the ground state for a range of values $(\omega_z, g^2)$. The number of photons serves as the order parameter, being zero in the ferromagnetic phase and finite in the superradiant phase, 
see Fig. \ref{fig: phase diagram QTFIM}(b). 
See Appendix $1$
for a detailed analysis of the phase transition point.
Further insight into this transition is obtained from the mean-field treatment, as discussed in the following sections.

\begin{figure}
\hfill{}\includegraphics[width=0.95\columnwidth]{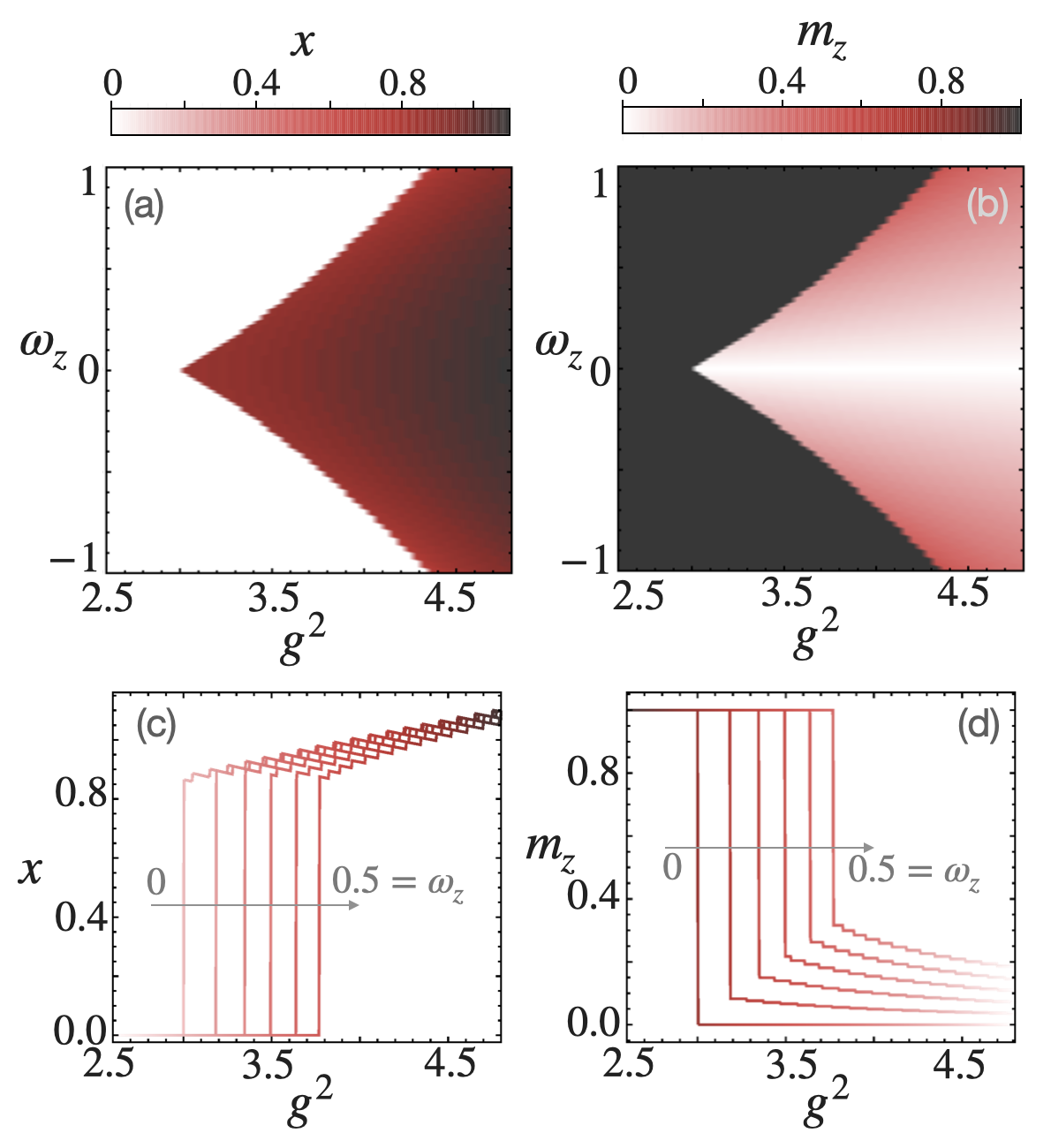}\hfill{}

\caption{The mean-field phase diagram for the (a) photon-coherent state $x$
and (b) magnetization $m_{z}$ as a function of the coupling strength
$g$ and the spin-excitation frequency $\omega_{z}$. The transition
is identified by the sudden change of colors, corresponding to the black line in Fig. 1(b). The lower plots show cuts
(fixed values of $\omega_{z}$) in the upper plots. Note that for
$\omega_{z}=0$, the magnetization $m_{z}$ is zero on the right side
of the transition. However, for $\left|\omega_{z}\right|>0$, the $m_{z}$
shifts from unity to a finite value; a similar, albeit slower, trend is observed for
$x$.}
\label{fig: mean-field}
\end{figure}

\paragraph*{Mean Field.}
The thermodynamic partition function can be approximated in the mean-field approach as
\begin{align}
Z & = \lim_{N_{s}\rightarrow\infty}\text{Tr}\left[\text{e}^{-\beta H_{\text{QTFIM}}}\right], \nonumber \\
& =\lim_{N_{s}\rightarrow\infty}\sqrt{\frac{N_{s}}{\beta\pi}}\int dx\exp\left[N_{s}f\left(x,m_{z}\right)\right],
\label{eq: Z in mean-field}
\end{align}
where
\begin{equation}
f\left(x,m_{z}\right) \equiv 
\left(-\beta\omega_{a}x^{2}+\ln\left[2\cosh\left(\beta\gamma/2\right)\right]-\beta J_{z}m_{z}^{2}\right),
\label{eq: f function}
\end{equation}
and $x\equiv\text{Re}\left[\alpha\right]/\sqrt{N_{s}}$ with $\alpha$ denoting the photonic coherent state $a\left|\alpha\right\rangle =\alpha\left|\alpha\right\rangle $.
We define $\gamma\equiv\sqrt{\left(2gx\right)^{2}+\left(4J_{z}m_{z}-\omega_{z}\right)^{2}}$,
representing the combined effects of interaction strengths and magnetization, with $m_{z}$ being the magnetization along the $z$ direction.
See Appendix $1$
for the complete derivation of eq. (\ref{eq: Z in mean-field}). 
In the thermodynamic limit, Laplace's method~\cite{PhysRevLett.93.083001,Gammelmark_2011} enables the determination of the optimal Gibbs-free energy by minimizing $f\left(x,m_{z}\right)$ with respect to $x$ and $m_{z}$. 
The results are depicted in Fig. \ref{fig: mean-field}. 
A clear phase transition is evident, indicated by a discontinuity in $x$ and $m_{z}$; 
this discontinuity corresponds to the black line in Fig. \ref{fig: phase diagram QTFIM}(b).

{\it Quench Dynamics and Confinement.}
Finite-system size effects tend to smooth the first-order mean-field transition curves into a continuous change of the order parameter and magnetization, see Appendix $1$.
This phenomenon allows for a few spin flips and the emergence of a few photon modes near the transition line. 
In this study, we have examined the quench dynamics of spin excitations in the QTFIM near transition.
Specifically, we start the system in the ferromagnetic phase and then abruptly shift it towards the transition line (this quench is represented in Fig. \ref{fig: phase diagram QTFIM}(b)) by increasing the spin-photon coupling $g$ and introducing the frequency splitting $\omega_z$.
In a competing fashion, the parameter $g$ facilitates spin-flips in the $z$ basis and photon creation, while $\omega_z$ promotes the alignment of all spins.

\begin{figure}
\hfill{}\includegraphics[width=0.95\columnwidth]{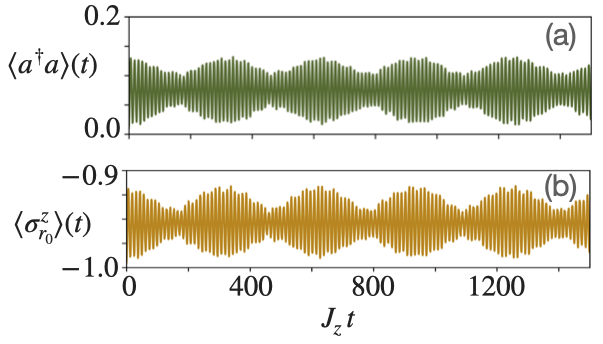}\hfill{}

\caption{Quench dynamics of the photon
number and $z$-magnetization, respectively, for $N_{p}=30$ for the photon and $N_{s}=14$ for the spin. We set $J_{z}=\omega_{a}$. 
The quench parameters are $\left(\omega_{z},g^{2}\right)=\left(0,2\right)\rightarrow\left(0.5,4\right)$
and are signaled in Fig. \ref{fig: phase diagram QTFIM}. The dynamics are characterized
by a rapid oscillation (with a period of $\Delta J_z \, t\approx10$) modulated by
a longer-period oscillation (with a period of $\Delta J_z \, t\approx300$).}
\label{fig: quench dynamics}
\end{figure}

\begin{figure}
\hfill{}\includegraphics[width=0.95\columnwidth]{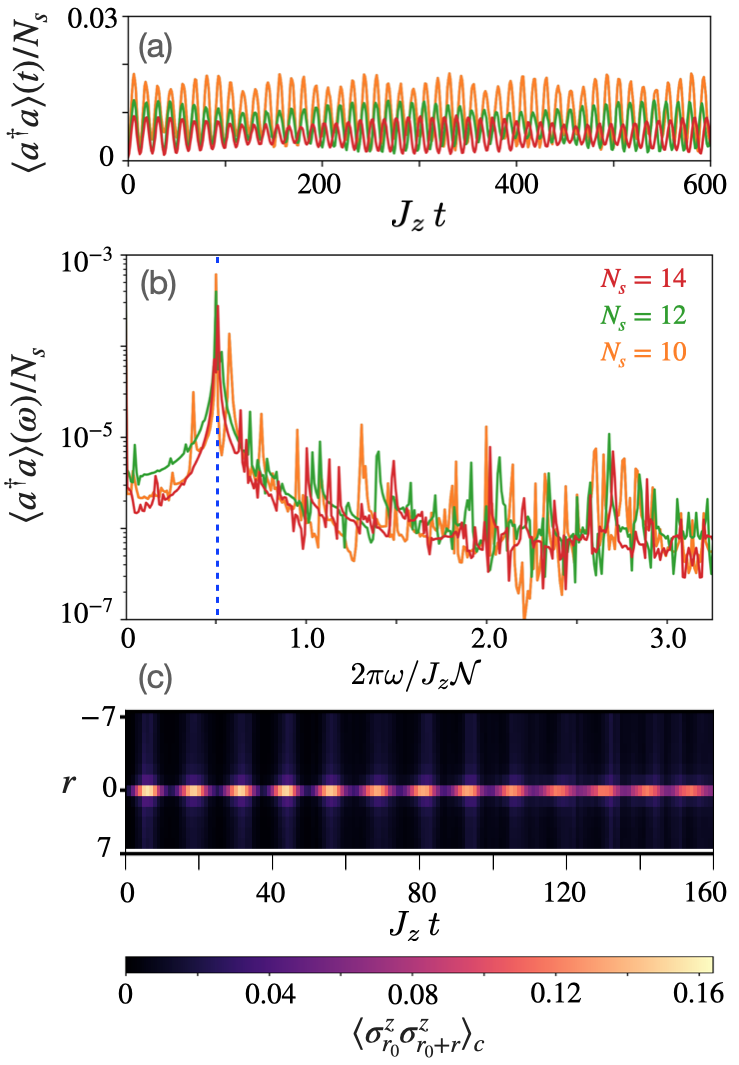}\hfill{}

\caption{(a) The same results as in Fig. \ref{fig: quench dynamics}(a) but for various system sizes.
Note that the fast-oscillation period is nearly independent of the
system size, while its modulation's period increases with the system
size. The legends follow panel (b), which displays their respective Fourier
transformation, with ${\cal N}=3000$. The highest peak, indicated by the
blue-vertical-dashed line, marks the frequency $\omega / J_z \approx 0.08 {\cal N}$ associated with
the fast oscillation. (c) The connected spin-spin correlation function during the quench dynamics. Observe that the correlations remain short-ranged
in space (at a distance $r$ between any two sites), indicating
the confinement of the excitations, and that the correlation approaches
zero every time the system nears the magnetic phase, with $\langle a^{\dagger}a\rangle=0$.
Periodic-boundary conditions were used for the exact diagonalization.}
\label{fig: quench dynamics correlations}
\end{figure}

Similar to the TFIM, small clusters of spin excitations tend to remain confined within small regions, and the dynamics is characterized by oscillations in the size of these clusters. 
%being that growth motivated by the strength $g$ and the shrink due to $\omega_z$. 
In the QTFIM, this dynamics can be tracked via the number of photons $\langle a^\dagger a\rangle (t)$ and the total magnetization $\langle \sigma_{r_0}^z \rangle (t)$ at time $t$, as shown in Fig. \ref{fig: quench dynamics}.
We observe a decrease in the magnetization followed by an increase in the photon number. Also, the oscillations exhibit two characteristic frequencies, i.e., a fast oscillation modulated by a slower one. The fast oscillation is nearly insensitive to the system size, while the slower one is strongly dependent, as illustrated in Fig. \ref{fig: quench dynamics correlations}. These frequencies can be determined by taking the Fourier transform
\begin{equation}
    \left\langle a^{\dagger}a\right\rangle \left(\omega_n \right)=\frac{1}{{\cal N}}\sum_{k}\text{e}^{-i \omega_n t_k }\left\langle a^{\dagger}a\right\rangle \left(t_{k}\right),
\end{equation}
where ${\cal N}$ is the number of terms in the sum, 
$\omega_n/J_z = n$, and $J_z t_k = 2\pi k/{\cal N}$. The indexes run over $n,k=1,2,\ldots , {\cal N}$.
The fast oscillation frequency is discerned from the highest peak in the Fourier transformation, as depicted in Fig. \ref{fig: quench dynamics correlations}(b). 
Confinement of spin excitations is evident by computing the connected correlation function (using periodic boundary conditions), 
$\left\langle \sigma_{r_{0}}^{z}\sigma_{r_{0}+r}^{z}\right\rangle _{c}\equiv\left\langle \sigma_{r_{0}}^{z}\sigma_{r_{0}+r}^{z}\right\rangle -\left\langle \sigma_{r_{0}}^{z}\right\rangle \left\langle \sigma_{r_{0}+r}^{z}\right\rangle$.
Fig. \ref{fig: quench dynamics correlations}(c) shows a synchronized increase of the correlations with photon creation over time. 

\begin{figure}
\hfill{}\includegraphics[width=0.95\columnwidth]{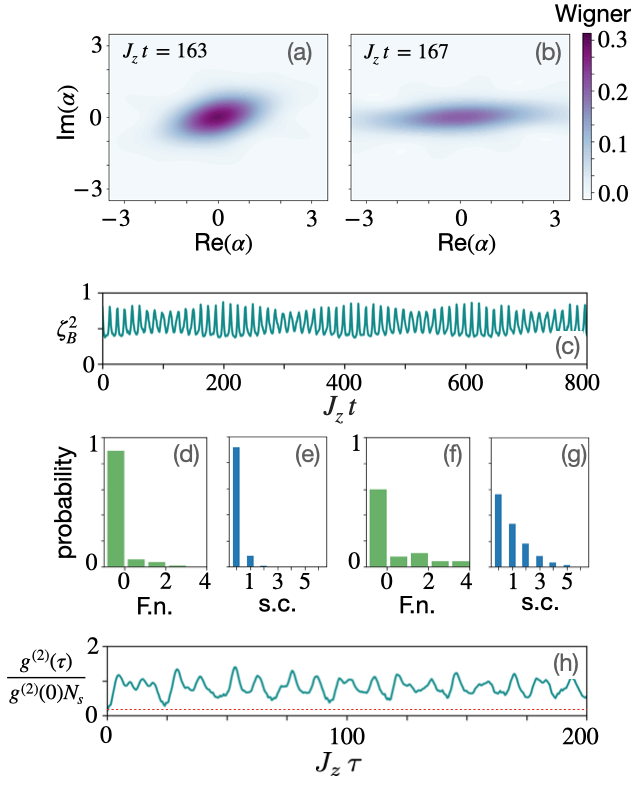}\hfill{}

\caption{Wigner function at $J_z \, t=163$ (a) and $J_z \, t=167$ (b) where the minimum and maximum photon excitation happen, respectively. 
The oscillation is accompanied by a squeezing in the photonic state.
The squeezed vacuum state is characterized by the squeezing factor $\zeta_{B}^{2}<1$ shown in (c).
For a better understanding of the photonic and magnetic states, we show the Fock number (F.n.) and the spin configuration (s.c.) in panels (d) and (e), respectively, 
for the same parameters as in (a). 
The (f) and (g) panels show analogous results but for the same parameters as in (b). 
Observe in panel (f) the pair-wise creation of photons ($0,2,4,\ldots$ photons), favoring over single photon creation.
Finally, the quench is characterized by a
correlation function $g^{\left(2\right)}\left(\tau\right)>g^{\left(2\right)}\left(0\right)$
at any instant $\tau$, as seen in (h), where the red-dashed line is $1/N_s$ for $N_s = 6$. }
\label{fig: Wigner function}
\end{figure}

{\it Characterization of the Photonic Field.}
We now turn our attention to analyzing the quench dynamics by examining the photonic properties.
We compute the dynamics of the Wigner function of the photonic field, initially in a vacuum state, which is a Gaussian distribution centered at zero.
We observe ``breathing'' oscillations of the Wigner function, which undergo oscillations and rotations in the phase space \cite{schleich2011quantum}, as illustrated in Figs. \ref{fig: Wigner function}(a) and \ref{fig: Wigner function}(b).

We monitor the degree of squeezing, measurable by the function $\zeta_B$ \cite{MA201189}
\begin{equation}
\zeta_{B}^{2}=1+2\left(\left\langle a^{\dagger}a\right\rangle -\left|\left\langle a\right\rangle \right|^{2}\right)-2\left|\left\langle a^{2}\right\rangle -\left\langle a\right\rangle ^{2}\right|.
\end{equation}
The condition for squeezing is $\zeta_{B}^{2} < 1$, as shown in Fig. \ref{fig: Wigner function}(c).
Additionally, we observe oscillations in the photon number distribution, attributed to interference in phase space between the ground and squeezed vacuum states \cite{Schleich:1987vp,Breitenbach:1997ug}, as shown in fig. \ref{fig: Wigner function}(f). 
The dynamics of the $g^{(2)}(\tau)$ correlation function
\cite{scully_zubairy_1997}
\begin{equation}
g^{\left(2\right)}\left(\tau\right)=\frac{\left\langle a^{\dagger}\left(0\right)a^{\dagger}\left(\tau\right)a\left(\tau\right)a\left(0\right)\right\rangle }{\left\langle a^{\dagger}\left(0\right)a\left(0\right)\right\rangle ^{2}},
\end{equation} 
is depicted in Fig. \ref{fig: Wigner function}(h). We note that $g^{(2)}(\tau)>g^{(2)}(0)$ for any time $\tau$. 

{\it Open System Dynamics.}
We express the master equation as $\partial_{t}\rho=-\frac{i}{\hbar}\left[H_{\text{QTFIM}},\rho\right]+{\cal L}\left[\rho\right]$ \cite{JOHANSSON20121760,JOHANSSON20131234}, incorporating the Lindbladian
\begin{align}
{\cal L}\left[\rho\right] & =\gamma\sum_{n}\left(\sigma_{n}^{-}\rho\sigma_{n}^{+}-\frac{1}{2}\left\{ \sigma_{n}^{\dagger}\sigma_{n}^{-},\rho\right\} \right)\nonumber \\
 & \quad+2\kappa\left[a\rho a^{\dagger}-\frac{1}{2}\left\{ a^{\dagger}a,\rho\right\} \right],
\end{align}
with $\kappa$ denoting the leakage rate and $\gamma$ the atomic-decay rate. 
The first term accounts for independent atomic decay ($\sigma^-_n$), while the second term represents photon leakage from the cavity ($a$) due to imperfect mirrors, characterized by rate $\kappa$.

\begin{figure}
\hfill{}\includegraphics[width=0.95\columnwidth]{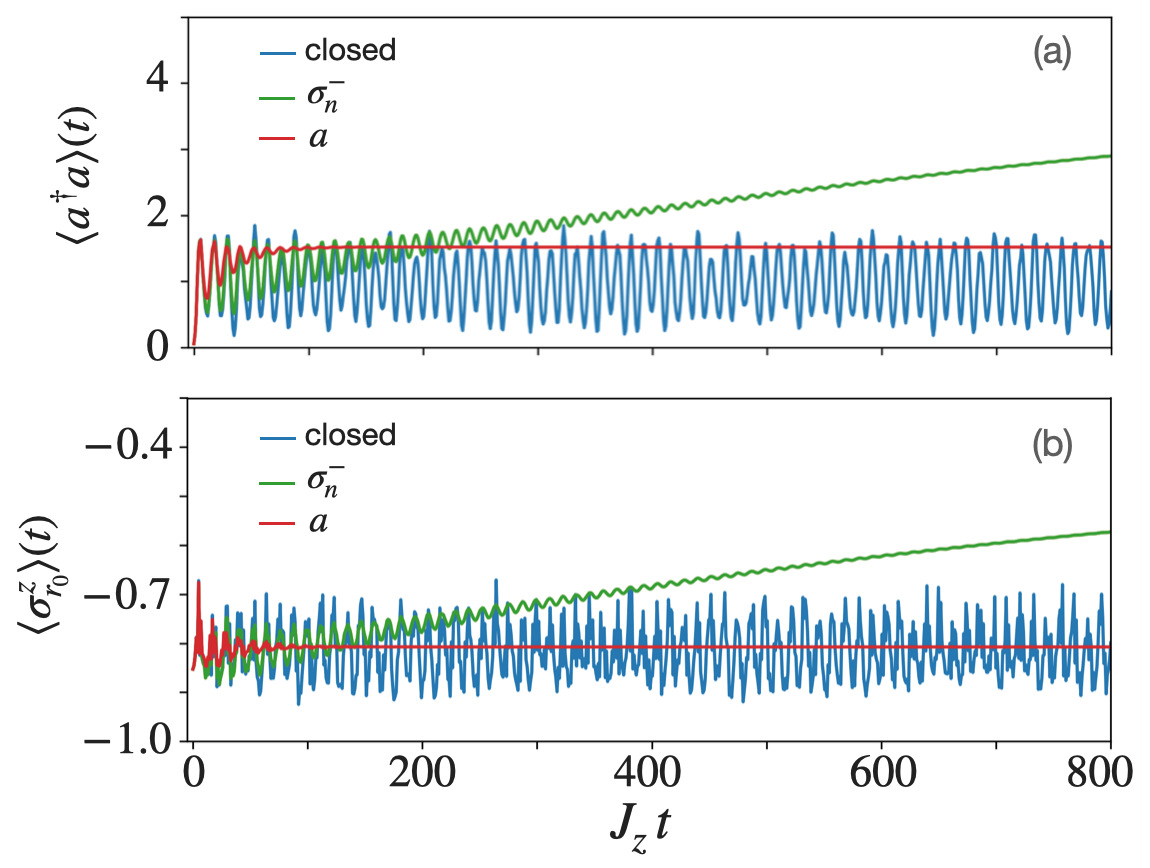}\hfill{}

\caption{(a) Photon number and (b) $z$-magnetization of an open system for
each one of the decay processes, described by the operators $\sigma_{n}^{-}$ and $a$. 
The results are juxtaposed with the quench dynamics of the closed
system (blue line), using the same parameters as in Fig. \ref{fig: phase diagram QTFIM}(b), with
$N_{s}=6$ and $N_{p}=20$. 
The plots present three independent simulations: $\gamma=0.01$, $\kappa=0$
(green), $\gamma=0$, $\kappa=0.01$ (red), and $\gamma=\kappa=0$ (blue).
}
\label{fig: open system}
\end{figure}

In Fig. \ref{fig: open system}, we examine these decay processes individually and compare the results with the closed system dynamics, using the same quench parameters as in Fig. \ref{fig: phase diagram QTFIM}(b).
The closed system ($\gamma = \kappa = 0$) exhibits oscillations in both photon number and magnetization, consistent with the confinement effect described earlier.
Interestingly, including cavity leakage in the time evolution reveals that the confinement-induced oscillations are suppressed. 
Nevertheless, the equilibrium state retains a finite photon number, indicating that the system remains in the superradiant phase.
Conversely, atomic decay exerts a less dramatic effect, allowing the oscillations to persist and resulting in a longer equilibration time. 

{\it Conclusions.}
This study represents a significant advancement in the understanding of confinement dynamics in quantum spin systems, particularly in the context of their interaction with cavity fields. We have explored the equilibrium phase diagram and the dynamics of confinement in such systems. Our work integrates the effective spin-spin interactions among Rydberg atoms with the Dicke model in a cavity-assisted setting, paving the way for novel experimental investigations in quantum many-body physics.

The feasibility of this model is further underscored by its compatibility with existing experimental platforms, and the theoretical model is shown to be realizable with a Rydberg-dressed chain of atoms coupled to a cavity field.
Finally, we highlight that the confinement effects unveiled in this research can be detected through optical readout, utilizing photon-cavity leakage, or via single-site imaging techniques. These methods are applicable to current experimental setups involving neutral-atom arrays, showcasing the broad applicability and potential impact of our findings in the field of quantum physics.

\begin{acknowledgments}
{\it Acknowledgments. --}
We acknowledge C. Dag, F. Liu, S. Ostermann, G. Pagano, and S. Yelin for useful discussions.
This work was supported by the Serrapilheira Institute 
(grant number Serra-1812-27802).
T.P. acknowledges the hospitality of the Physics Department of UFRN and the Serrapilheira Institute for support.
We thank the High-Performance Computing Center (NPAD) at UFRN for providing computational resources.
\end{acknowledgments}

% -------------------------------

\bibliographystyle{apsrev4-1}
\bibliography{refs}

% -------------------------------

\clearpage
\pagebreak
\widetext
\begin{center}
\textbf{\large Supplemental Material:\\[5pt] Confined Meson Excitations in Rydberg-Atom Arrays Coupled to a Cavity Field\\[10pt]}
\textrm{Tharnier O. Puel$^{1,2}$ and Tommaso Macr\`{i}$^{3,2}$\\[3pt]
$^{1}$Department of Physics and Astronomy, University of Iowa, Iowa City, Iowa 52242, USA\\
$^{2}$Departamento de F\'{i}sica Te\'{o}rica e Experimental,  Universidade Federal do Rio Grande do Norte, Campus Universit\'{a}rio, Lagoa Nova, Natal-RN 59078-970, Brazil\\
$^{3}$ITAMP, Harvard-Smithsonian Center for Astrophysics, Cambridge, Massachusetts 02138, USA\\
}
\end{center}
%%%%%%%%%% Merge with supplemental materials %%%%%%%%%%
%%%%%%%%%% Prefix a "S" to all equations, figures, tables and reset the counter %%%%%%%%%%
\renewcommand{\theequation}{S\arabic{equation}}
\renewcommand{\thefigure}{S\arabic{figure}}
%%%%%%%%%% Prefix a "S" to all equations, figures, tables and reset the counter %%%%%%%%%%

\appendix

\subsection{1. Ground-state Phase Diagram}
\label{app: phase diagram}

\begin{figure}[h!]
\hfill{}\includegraphics[width=0.75\columnwidth]{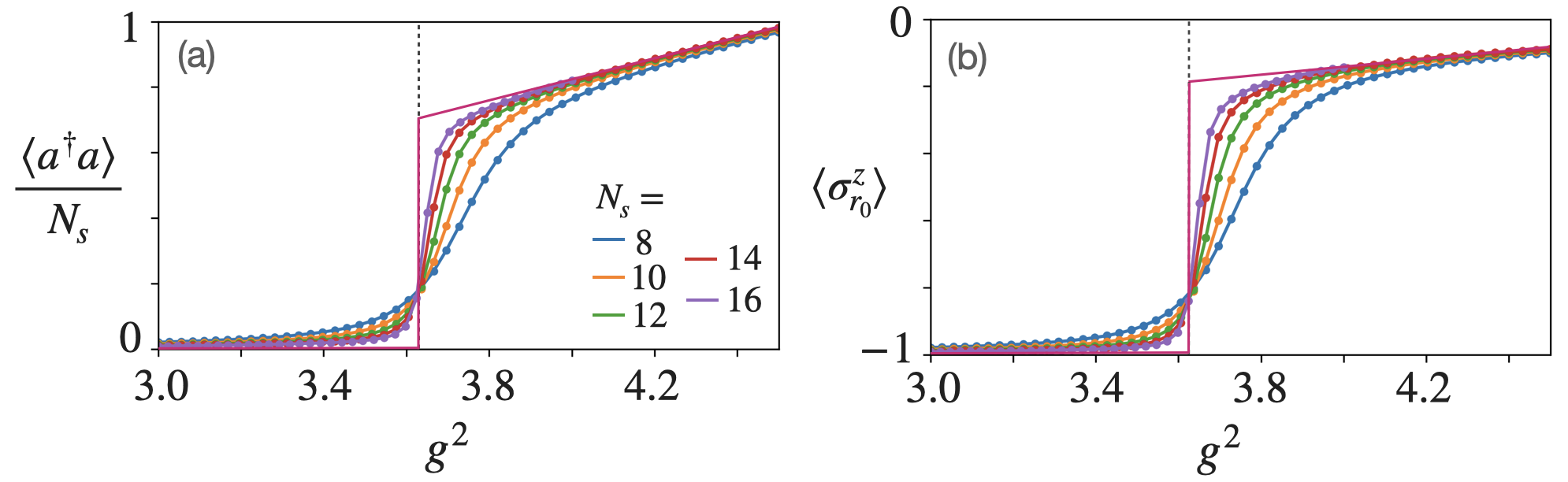}\hfill{}
\caption{(a) Photon number rescaled by the number of spins and (b) $z$-magnetization as a function of the
coupling strength $g$. This illustrates the phase transition for $\omega_{z}=0.1$,
where the finite-size analysis enables us to identify $g_{c}^{2}\left(\omega_{z}=0.1\right)=3.62$,
as indicated by the vertical-dashed line. The solid lines serve as a guide to the eye.}
\label{fig: critical point}
\end{figure}

In Fig. \ref{fig: critical point}, we present an example of the scaling analysis with the system size. It becomes clear that, for $\omega_z \neq 0$, there is a phase transition in the thermodynamic limit ($N_s \rightarrow \infty$).

%%%% ------------------------------------------------------------------------------------   

\subsection*{Partition Function}

We restate Eqs. (\ref{eq:H Ising}) and (\ref{eq:H Dicke}) here for convenience,

\begin{align}
H_{\text{QTFIM}}= 
& -J_{z}\sum_{n}\sigma_{n}^{z}\sigma_{n+1}^{z} \nonumber \\
& + \omega_{z}S_{z}+\frac{g}{\sqrt{N_{s}}}\left(a^{\dagger}+a\right)S_{x}+\omega_{a}a^{\dagger}a,\label{eq: QTFIM appendix}
\end{align}
where we have defined $S_{x/z}\equiv\sum_{n}\sigma_{n}^{x/z}/2$.
Under the assumption that $b\equiv a/\sqrt{N_{s}}$ and $b^{\dagger}\equiv a^{\dagger}/\sqrt{N_{s}}$
exist for $N_{s}\rightarrow\infty$, such that their commutator $\left[b,b^{\dagger}\right]=1/N_{s}$
vanishes in the thermodynamic limit. The partition function is given by
\begin{align}
Z & =\lim_{N_{s}\rightarrow\infty}\text{Tr}\left[\text{e}^{-\beta H_{\text{QTFIM}}}\right]\nonumber \\
 & =\text{Tr}\left[\lim_{R\rightarrow\infty}\lim_{N_{s}\rightarrow\infty}\sum_{r=0}^{R}\frac{\left(-\beta H_{\text{QTFIM}}\right)^{r}}{r!}\right]\nonumber \\
 & =\text{Tr}\left[\text{e}^{-\beta H_{a}}\text{e}^{-\beta H'}\right],
\end{align}
where $H_{a}\equiv\omega_{a}a^{\dagger}a$, and thus $H_{\text{QTFIM}}=H'+H_{a}$.
Note in the second line that we have assumed that the limits can
be interchanged, therefore in the thermodynamic limit the exponentials
decouple as shown in the last line. In order to proceed with
the calculations, it is convenient to utilize the coherent states, $a\left|\alpha\right\rangle =\alpha\left|\alpha\right\rangle $,
$\left\langle \alpha\right|a^{\dagger}=\left\langle \alpha\right|\alpha^{*}$,
and $\int\frac{d^{2}\alpha}{\pi}\left|\alpha\right\rangle \left\langle \alpha\right|=1$,
leading to
\begin{align}
Z & =\text{Tr}\left[\text{e}^{-\beta H_{a}}\text{e}^{-\beta H'}\right]\nonumber \\
 & =\int\frac{d^{2}\alpha}{\pi}\text{e}^{-\beta\left\langle \alpha\right|H_{a}\left|\alpha\right\rangle }\text{Tr}_{\text{spin}}\left[\text{e}^{-\beta\left\langle \alpha\right|H'\left|\alpha\right\rangle }\right]\nonumber \\
 & =\int\frac{d^{2}\alpha}{\pi}\text{e}^{-\beta\omega_{a}\left|\alpha\right|^{2}}\text{Tr}_{\text{spin}} {\bigg [ } \nonumber \\
 &
 \left.\exp\left[-\beta\left(-J_{z}\sum_{n}\sigma_{n}^{z}\sigma_{n+1}^{z}+\omega_{z}S_{z}+\Omega_{x}S_{x}\right)\right]\right],
\end{align}
where $\Omega_{x}\equiv2g\text{Re}\left[\alpha\right]/\sqrt{N_{s}}$.

% --------------------------------------
\paragraph{Mean-field Approach.}

We now apply the mean-field approach and identify
\begin{align}
\sum_{n}\sigma_{n}^{z}\sigma_{n+1}^{z}=
&\sum_{n}\left[m_{z}\left(\sigma_{n}^{z}+\sigma_{n+1}^{z}\right)-m_{z}^{2}\right] \approx 2m_{z}\sum_{n}\sigma_{n}^{z}-N_{s}m_{z}^{2},
\end{align}
where $m_{z}\equiv\left\langle \sigma_{n}^{z}\right\rangle $ and
we have considered a translation-invariant system. The Hamiltonian
$H'$ then becomes
\begin{align}
H' & =-J_{z}\sum_{n}\sigma_{n}^{z}\sigma_{n+1}^{z}+\omega_{z}S_{z}+\Omega_{x}S_{x}\nonumber \\
 & \approx-\omega_{z,\text{eff}}S_{z}+\Omega_{x}S_{x}+\text{const.},
\end{align}
where $\omega_{z,\text{eff}}=4J_{z}m_{z}-\omega_{z}$ and $\text{const.}=N_{s}J_{z}m_{z}^{2}$.
To facilitate further analysis, it is useful to rotate the spin axis of quantization along the
$y$-direction, resulting in the spin operators transforming as follows:
\begin{align}
\text{e}^{i\phi\sigma^{y}}\left(-\omega_{z,\text{eff}}\sigma_{n}^{z}+\Omega_{x}\sigma_{n}^{x}\right) & =\left(\Omega_{x}\cos\phi+\omega_{z,\text{eff}}\sin\phi\right)\sigma_{n}^{x}\nonumber \\
 & \quad+\left(\Omega_{x}\sin\phi-\omega_{z,\text{eff}}\cos\phi\right)\sigma_{n}^{z}.
\end{align}
By choosing $\phi=\arctan\left[-\Omega_{x}/\omega_{z,\text{eff}}\right]$, we obtain $\sin\phi=-\Omega_{x}/c$ and $\cos\phi=\omega_{z,\text{eff}}/c$,
with $c^{2}=\Omega_{x}^{2}+\omega_{z,\text{eff}}^{2}$, yielding $H'_{n}\rightarrow -\gamma\sigma_{n}^{z}+\text{const.}$, with $\gamma = \sqrt{ \Omega_{x}^{2}+\omega_{z,\text{eff}}^{2} }$.

% --------------------------------------
\paragraph*{Back to the Partition Function.}

Finally, the partition function simplifies to
\begin{align}
Z & =\int\frac{d^{2}\alpha}{\pi}\text{e}^{-\beta\omega_{a}\left|\alpha\right|^{2}}\text{Tr}_{\text{spin}}\left[\exp\left[-\beta\left(-\gamma S_{z}+\text{const.}\right)\right]\right]\nonumber \\
 & =\int\frac{d^{2}\alpha}{\pi}\text{e}^{-\beta\omega_{a}\left|\alpha\right|^{2}}\left[2\cosh\left(\beta\gamma/2\right)\right]^{N_{s}}\text{e}^{-\beta\text{const.}}.
\end{align}
Resolving the integral over coherent states, $d^{2}\alpha=d\text{Re}\left[\alpha\right]d\text{Im}\left[\alpha\right]$,
and defining $x\equiv\text{Re}\left[\alpha\right]/\sqrt{N_{s}}$,
leads us to the final expression
\begin{align}
Z =
&\sqrt{\frac{N_s}{\beta\pi}}\int dx\exp \big [\left. N_{s}\left(-\beta\omega_{a}x^{2}+\ln\left[2\cosh\left(\beta\gamma/2\right)\right]-\beta J_{z}m_{z}^{2}\right)\right],\label{eq: partition function integral form}
\end{align}
which corresponds to Eq. (\ref{eq: Z in mean-field}) presented in the main text.

% --------------------------------------
\subsection*{Self-consistent Magnetization}

Due to the integration over $x$ in Eq. (\ref{eq: partition function integral form}),
the Gibbs energy, $G=-\frac{1}{\beta}\ln\left[Z\right]$, is not
easily extracted. However, the thermodynamic limit ($N_s \rightarrow \infty$) can be analyzed by applying Laplace's method~\cite{PhysRevLett.93.083001,Gammelmark_2011}, which states that if the function $f\left(x\right)$ within the integral $F\left(N\right)=\int_{a}^{b}\text{e}^{-Nf\left(x\right)}dx$
has a minimum at $x_{0}$, then 
\begin{equation}
F\left(N\right)=\sqrt{\frac{2\pi}{f''\left(x_{0}\right)N}}\text{e}^{-Nf\left(x_{0}\right)},\quad N\rightarrow\infty.
\end{equation}
Therefore, the partition function can be approximated by
\begin{align}
Z & \approx\sqrt{\frac{2}{\beta f''\left(x\right)}}\text{e}^{N_{s}\left(-\beta\omega_{a}x^{2}+\ln\left[2\cosh\left(\beta\gamma/2\right)\right]-\beta J_{z}m_{z}^{2}\right)},\; N_{s}\rightarrow\infty.
\end{align}
Now, the Gibbs free energy can be computed in the usual form
\begin{align}
G  =
&-\frac{N_{s}}{\beta}\left(-\beta\omega_{a}x^{2}+\ln\left[2\cosh\left(\beta\gamma/2\right)\right]-\beta J_{z}m_{z}^{2}\right) +\frac{1}{2\beta}\ln\left[f''\left(x\right)\right]+\text{constant},
\end{align}
where
\begin{align}
f''\left(x\right) & =\frac{\partial^{2}}{\partial^{2}x}\left(-\beta\omega_{a}x^{2}+\ln\left[2\cosh\left(\beta\gamma/2\right)\right]-\beta J_{z}m_{z}^{2}\right) =-2\beta\omega_{a}+\frac{\partial^{2}}{\partial^{2}x}\ln\left[2\cosh\left(\beta\gamma/2\right)\right],
\end{align}
with $\gamma=\sqrt{\omega_{z,\text{eff}}^{2}+\left(2gx\right)^{2}}$.
Consequently, the magnetization in the $z$-direction is given by
\begin{align}
m_{z} & =  -\frac{1}{N_{s}}\frac{\partial G}{\partial\omega_{z,\text{eff}}}\nonumber \\
 & =  \frac{1}{\beta}\frac{\partial\ln\left[2\cosh\left(\beta\gamma/2\right)\right]}{\partial\omega_{z,\text{eff}}} -\frac{1}{2N_{s}\beta}\frac{\partial}{\partial\omega_{z,\text{eff}}}\frac{\partial^{2}}{\partial^{2}x}\ln\left[2\cosh\left(\beta\gamma/2\right)\right].
\end{align}
In the thermodynamic limit, $N_{s}\rightarrow\infty$, the second
term on the right-hand side of the above expression becomes negligible,
and the magnetization simplifies to
\begin{align}
m_{z} & =\frac{1}{2}\tanh\left(\beta\gamma/2\right)\frac{\omega_{z,\text{eff}}}{\gamma}.
\end{align}
This result depends on $x$ and thus must be determined
self-consistently with Eq. (\ref{eq: f function}) in the main text to minimize
the free energy.

% --------------------------------------
\section{Two-kink Model}

We define the quantum transverse field Ising model (QTFIM) as
\begin{align}
H & \equiv H_{z}+H_{x},\\
H_{z} & \equiv-J_{z}\sum_{n}\sigma_{n}^{z}\sigma_{n+1}^{z}+\omega_{z}S_{z}+\omega_{a}a^{\dagger}a,\\
H_{x} & \equiv\frac{g}{\sqrt{N_{s}}}\left(a^{\dagger}+a\right)S_{x},
\end{align}
where $S_{\alpha}=\sum_{i}\sigma_{i}^{\alpha}/2$ and $N_{s}$ is the total
number of spins.

\paragraph*{Two-kink States. }

Consider a projected Hilbert space, where the Hamiltonian is spanned
by the states with $n$ spins down, represented as
\begin{equation}
\left|j,n,n_{p}\right\rangle \equiv\left|\cdots\uparrow\uparrow\downarrow_{j}\downarrow\cdots\downarrow\downarrow_{\left(j+n-1\right)}\uparrow\uparrow\cdots\right\rangle \left|n_{p}\right\rangle ,
\end{equation}
with $a\left|n_{p}\right\rangle =\sqrt{n_{p}}\left|n_{p}-1\right\rangle$ and $a^{\dagger}\left|n_{p}\right\rangle =\sqrt{n_{p}+1}\left|n_{p}+1\right\rangle $.
Note that there are only two domain walls: between sites $j$ and
$\left(j-1\right)$, as well as sites $\left(j+n-1\right)$ and $\left(j+n\right)$.
In this basis, not more nor less domain walls are allowed. In the following, we will evaluate
how the Hamiltonian acts on that subspace. 
We start our analysis by evaluating
the transverse field
\begin{align}
H_{x}\left|j,n,n_{p}\right\rangle = & \frac{g}{2\sqrt{N_{s}}}\left(a^{\dagger}+a\right)\sum_{i}\sigma_{i}^{x}\left|j,n,n_{p}\right\rangle \nonumber \\
= & \frac{g}{2\sqrt{N_{s}}}\left(\left|j-1,n+1\right\rangle +\left|j+1,n-1\right\rangle \right.\nonumber \\
 & \left.+\left|j,n-1\right\rangle +\left|j,n+1\right\rangle \right)\nonumber \\
 & \times\left(\sqrt{n_{p}+1}\left|n_{p}+1\right\rangle +\sqrt{n_{p}}\left|n_{p}-1\right\rangle \right) .
\end{align}
Note that, when $\sigma_{j}^{x}$ flips the $j$-th spin (flips
it up), the cluster of down-spins decreases by one, i.e., $n\rightarrow\left(n-1\right)$.
On the other hand, when $\sigma_{j-1}^{x}$ flips the $\left(j-1\right)$-th
spin (flips it down), the cluster increases by one, i.e., $n\rightarrow\left(n+1\right)$.

The total energy depends on the number of spins down, thus it
is convenient to measure this energy relative to the ground-state energy
$E_{0}\equiv\left\langle \uparrow\uparrow\cdots\uparrow\uparrow,n_{p}=0\right|H_{z}\left|\uparrow\uparrow\cdots\uparrow\uparrow,n_{p}=0\right\rangle $.
Having established this, we compute $E_{0}=\left(-J_{z}N+\frac{\omega_{z}}{2}N\right)$
and define
\begin{align}
V\left(n,n_{p}\right) & \equiv\left\langle j,n,n_{p}\right|H_{z}\left|j,n,n_{p}\right\rangle -E_{0}\nonumber \\
 & =J_{z}4-\omega_{z}n+\omega_{a}n_{p}.
\end{align}

To compute
\begin{equation}
\tilde{H}=\sum_{n_{p},m_{p}}\sum_{j,n}\sum_{l,m}\left\langle j,n,n_{p}\right|H-E_{0}\left|l,m,m_{p}\right\rangle \left|j,n,n_{p}\right\rangle \left\langle l,m,m_{p}\right|,
\end{equation}
we first perform the Fourier transform, from the position $j$
to a momentum $k$, defined as
\begin{align}
\left|k,n,n_{p}\right\rangle  & =\frac{1}{\sqrt{L}}\sum_{j=1}^{L}\text{e}^{-ik\left(j+n/2\right)}\left|j,n,n_{p}\right\rangle ,\\
\left|j,n,n_{p}\right\rangle  & =\frac{1}{\sqrt{L}}\sum_{k}\text{e}^{ik\left(j+n/2\right)}\left|k,n,n_{p}\right\rangle .
\end{align}
The resulting Hamiltonian is
\begin{align}
\tilde{H} & =\sum_{k}\sum_{n_{p}}\sum_{n}V\left(n,n_{p}\right)\left|k,n,n_{p}\right\rangle \left\langle k,n,n_{p}\right|\nonumber \\
 & \qquad+\frac{g}{\sqrt{N_{s}}}\cos\left(k/2\right)\left(\left|k,n\right\rangle \left\langle k,n-1\right|+\left|k,n\right\rangle \left\langle k,n+1\right|\right)\nonumber \\
 & \qquad\times\left(\sqrt{n_{p}}\left|n_{p}\right\rangle \left\langle n_{p}-1\right|+\sqrt{n_{p}+1}\left|n_{p}\right\rangle \left\langle n_{p}+1\right|\right) ,
\end{align}
where we have shown that $V\left(n,n_{p}\right)=J_{z}4-\omega_{z}n+\omega_{a}n_{p}$.

\section{2. Details on the Implementation}

Our theoretical proposal is based on the possibility of implementing the Ising-Dicke model with a Rydberg-dressed chain of atoms coupled to a cavity field (see also Ref. \cite{SciPostPhys.1.1.004}). 
The Hamiltonian of the system is expressed as
\begin{equation}
H=H_{\text{cav}}+H_{\text{atom}}+H_{\text{pump}}\left(t\right)+H_{\text{atom-light}}+H_{\text{atom-atom}},
\label{eq: Ising-Dicke implementation}
\end{equation}
where
\begin{align}
H_{\text{cav}}= &\, \omega_{c}\, a^{\dagger}a,\\
H_{\text{atoms}}= & \sum_{\ell=1}^{N}\omega_{d}\left|d\right\rangle _{\ell}\left\langle d\right|+\omega_{e}\left|e\right\rangle _{\ell}\left\langle e\right|\nonumber \\
 & +\omega_{\text{Ryd}}\left|\text{Ryd}\right\rangle _{\ell}\left\langle \text{Ryd}\right|+\omega_{1}\left|1\right\rangle _{\ell}\left\langle 1\right|\\
H_{\text{pump}}\left(t\right)= & \sum_{\ell=1}^{N}\frac{\Omega_{e}}{2}e^{-i\omega_{\Delta e}t}\left|e\right\rangle _{\ell}\left\langle 1\right|+\frac{\Omega_{d}}{2}e^{-i\omega_{\Delta d}t}\left|d\right\rangle _{\ell}\left\langle 0\right|\nonumber \\
 & +\frac{\Omega_{\text{Ryd}}}{2}e^{-i\omega_{\Delta r}t}\left|\text{Ryd}\right\rangle _{\ell}\left\langle 1\right|+\text{H.c.},\\
H_{\text{atom-light}}= & \sum_{\ell=1}^{N}\left(g_{d}\left|d\right\rangle _{\ell}\left\langle 1\right|+g_{e}\left|e\right\rangle _{\ell}\left\langle 0\right|\right)a+\text{H.c.},\\
H_{\text{atom-atom}}= & \sum_{\left\langle lm\right\rangle }V_{lm}\left(\left|\text{Ryd}\right\rangle _{\ell}\left\langle \text{Ryd}\right|\right)\left(\left|\text{Ryd}\right\rangle _{m}\left\langle \text{Ryd}\right|\right).
\end{align}
The frequencies $\omega_{d},\omega_{e},\omega_{\text{Ryd}},\omega_{1}$
correspond to the atomic levels labelled by the sequence $\left|d\right\rangle ,\left|e\right\rangle ,\left|\text{Ryd}\right\rangle ,\left|1\right\rangle $,
and are measured relative to the atomic level $\left|0\right\rangle $.
Similarly, the frequencies $\omega_{\Delta d},\omega_{\Delta e},\omega_{\Delta r}$
pertain to laser frequencies of the pump-terms. Here, $\omega_{c}$
denotes the bare cavity resonance. We have assumed homogeneous pumping
of the atoms from the side $\Omega_{d;\ell}\approx\Omega_{d}$ and
$\Omega_{e;\ell}\approx\Omega_{e}$, as well as homogeneous coupling
of the light field to the atoms $g_{d;\ell}\approx g_{d}$ and $g_{e;\ell}\approx g_{e}$.

\begin{figure}
\hfill{}\includegraphics[width=0.35\columnwidth]{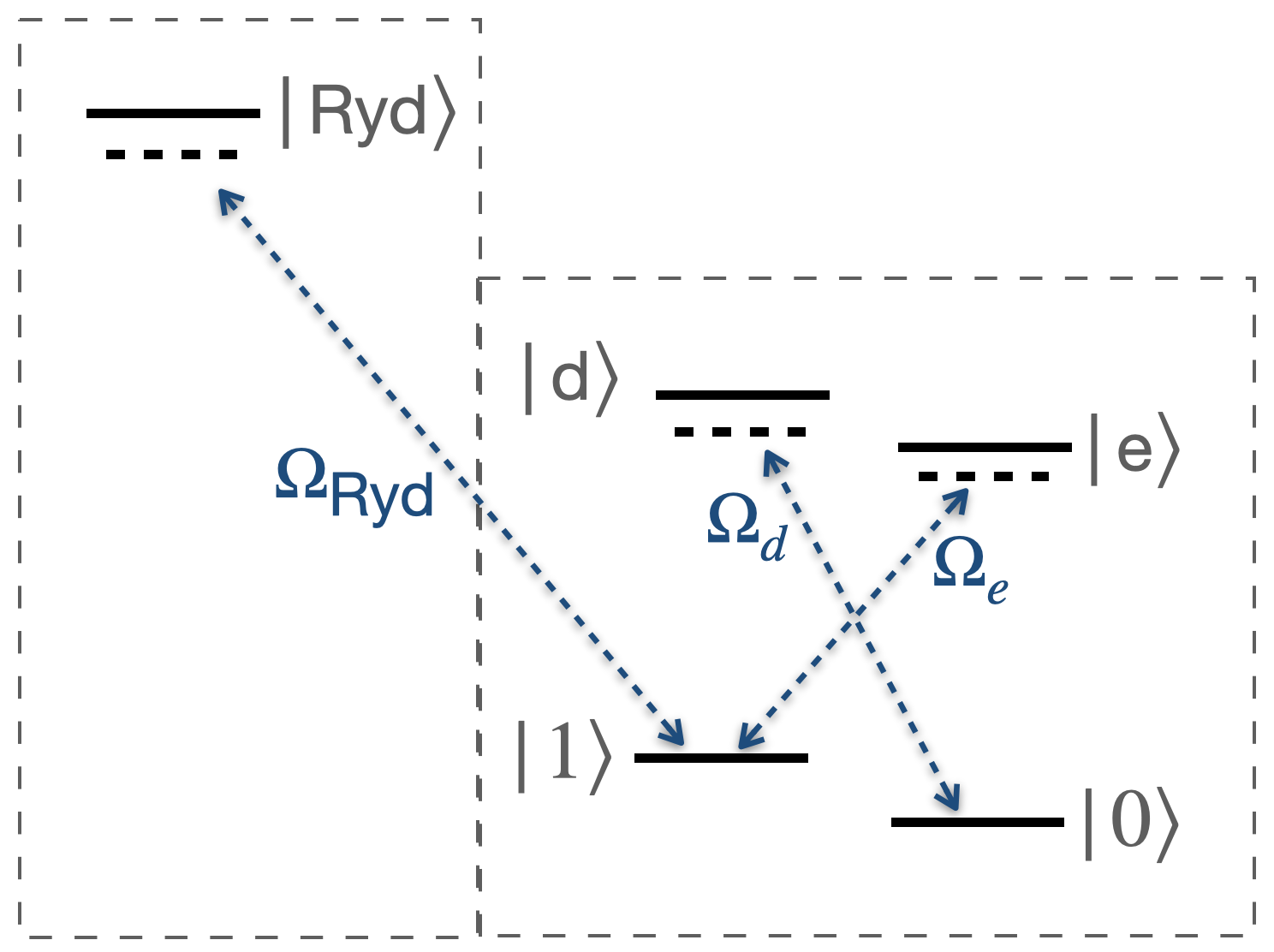}\hfill{}
\caption{Level scheme for the simulation of the Dicke-model via cavity-assisted Raman transitions and the Rydberg state that generates effective exchange interaction between neighboring atoms.}
\label{fig: energy levels}
\end{figure}

\subsection{Effective Hamiltonian}

After applying the rotating wave approximation to Eq. (\ref{eq: Ising-Dicke implementation}), one derives the effective Hamiltonian as described in Ref. \cite{SciPostPhys.1.1.004}
\begin{align}
\tilde{H}= & a^{\dagger}a\left[\frac{N}{2}\left(\frac{g_{e}^{2}}{\Delta_{e}}+\frac{g_{d}^{2}}{\Delta_{d}}\right)+\tilde{\omega}_{a}\right]\nonumber \\
+ & \sum_{\ell=1}^{N}\frac{\sigma_{\ell}^{z}}{2}\left[\left(\frac{\Omega_{d}^{2}}{4\Delta_{d}}-\frac{\Omega_{e}^{2}}{4\Delta_{e}}\right)+\Delta_{1}-\frac{\Omega_{\text{Ryd}}^{2}}{4\Delta_{\text{Ryd}}}+\frac{1}{2}\sum_{m;m\neq\ell}^{N}V_{m\ell}^{\text{eff}}\right]\nonumber \\
 & +\sum_{\ell=1}^{N}\sigma_{\ell}^{z}\frac{1}{2}\frac{1}{N}\left(\frac{g_{d}^{2}}{\Delta_{d}}-\frac{g_{e}^{2}}{\Delta_{e}}\right)a^{\dagger}a\nonumber \\
 & +\sum_{\ell=1}^{N}\left[\frac{\lambda_{d}}{\sqrt{N}}\left(\sigma_{\ell}^{+}a^{\dagger}+\sigma_{\ell}^{-}a\right)+\frac{\lambda_{e}}{\sqrt{N}}\left(\sigma_{\ell}^{+}a+\sigma_{\ell}^{-}a^{\dagger}\right)\right]\nonumber \\
 & +\frac{1}{2}\sum_{i\neq j}V_{ij}^{\text{eff}}\frac{\sigma_{i}^{z}}{2}\frac{\sigma_{j}^{z}}{2},\label{eq: eff Hamiltonian}
\end{align}
where the parameters are defined as follows:
\begin{align}
\omega'_{1} & =\left(\omega_{\Delta d}-\omega_{\Delta e}\right)/2,\\
\tilde{\omega}_{a} & =\omega_{c}-\left(\omega_{\Delta d}-\omega'_{1}\right),\\
\Delta_{\text{Ryd}} & =-\left[\omega_{\text{Ryd}}-\left(\omega_{\Delta r}+\omega'_{1}\right)\right],\\
\Delta_{1} & =\omega_{1}-\omega'_{1},\\
\Delta_{d} & =\omega_{d}-\omega_{\Delta d},\\
\Delta_{e} & =\omega_{e}-\omega_{\Delta e}.
\end{align}
The effective spin-photon coupling and spin-spin interaction strength are given respectively by
\begin{equation}
\lambda_{d/e}=\sqrt{N}\frac{g_{d/e}\Omega_{d/e}}{2\Delta_{d/e}},\qquad V_{ij}^{\text{eff}}=\left(\frac{\Omega_{\text{Ryd}}}{2\Delta_{\text{Ryd}}}\right)^{4}\frac{C_{6}}{r_{ij}^{6}+R_{c}^{6}},\label{eq: lambdas and Veff}
\end{equation}
where $R_{c}^{6}=C_{6}/\left(2\hbar\left|\Delta_{\text{Ryd}}\right|\right)$
is defined as the critical radius obtained by imposing $V\left(R_{c}\right)\equiv2\Delta_{\text{Ryd}}$.
$r_{ij}$ denotes the distance between Rydberg
atoms at positions $i$ and $j$. $C_{6}$ describes the Van der Waals interaction strength.

% ----------------------------------------
\subsection{Comparison with the Ising-Dicke Hamiltonian}

The quantized transverse field Ising model (QTFIM, or just Ising-Dicke Hamiltonian) as introduced in the main text reads
\begin{align}
H_{\text{QTFMI}}= & \omega_{a}a^{\dagger}a+\omega_{z}\sum_{\ell}\frac{\sigma_{\ell}^{z}}{2}\nonumber \\
 & +\frac{g}{2\sqrt{N}}\left(a^{\dagger}+a\right)\sum_{\ell}\left(\sigma_{\ell}^{+}+\sigma_{\ell}^{-}\right)\nonumber \\
 & -4J_{z}\sum_{i}\frac{\sigma_{i}^{z}}{2}\frac{\sigma_{i+1}^{z}}{2}.
 \label{app: QTFMI}
\end{align}
By directly comparing $H_{\text{QTFMI}}$ in Eq. (\ref{app: QTFMI}) with the effective
Hamiltonian in Eq. (\ref{eq: eff Hamiltonian}), we can correlate the
coefficients in each term. First, we consider only nearest-neighbor hoppings,
which renders the $J_{z}$ a constant.
Therefore, we have
\begin{equation}
V_{ij}^{\text{eff}}=\left(\frac{\Omega_{\text{Ryd}}}{2\Delta_{\text{Ryd}}}\right)^{4}\frac{C_{6}}{r_{ij}^{6}+R_{c}^{6}}=-8J_{z}\delta_{j,\left(i+1\right)},\qquad\forall\;i.
\end{equation}
Second, the $H_{\text{QTFMI}}$ does not include couplings of the type
$\sigma_{\ell}^{z}a^{\dagger}a$, which necessitates
\begin{equation}
\frac{1}{N}\left(\frac{g_{d}^{2}}{\Delta_{d}}-\frac{g_{e}^{2}}{\Delta_{e}}\right)\rightarrow0. \label{delta paper}
\end{equation}
This condition can be met by tuning the parameters $g_{d}^{2}/\Delta_{d}=g_{e}^{2}/\Delta_{e}\equiv \kappa$. 
Third, under the condition $\lambda_{d}=\lambda_{e}\equiv\lambda$, we derive the relation $\lambda=g/2$ between the two equations (note that $g$ should not be confused with $g_d$ or $g_e$).
Finally, it is straightforward to see that
\begin{align}
\omega_{a} & =\left[\frac{N}{2}\kappa+\omega_{c}-\frac{1}{2}\left(\omega_{\Delta d}+\omega_{\Delta e}\right)\right],
\label{omega_z correspondence 1}
\\
\omega_{z} & =\left[\Delta_{1}-\frac{\Omega_{\text{Ryd}}^{2}}{4\Delta_{\text{Ryd}}}-4J_{z}\right]. 
\label{omega_z correspondence 2}
\end{align}

\subsection{Range of Parameters for Experimental Realization}

In this final section, we discuss realistic parameters for implementing our model in an experiment. We begin by analyzing the Dicke model through Raman transitions within an optical cavity, as in Ref. \cite{SciPostPhys.1.1.004}. The proposed realization uses $^{87}$Rb $D_1$ transitions. Thus, we consider the hyperfine states $|5^2 S_{1/2}, F=1\rangle \equiv |0\rangle$ and $|5^2 S_{1/2}, F=2\rangle \equiv |1\rangle$ as the ground states, and $|5^2 P_{1/2}, F=1\rangle \equiv |e\rangle$ and $|5^2 P_{1/2}, F=2\rangle \equiv |d\rangle$ as the excited states. The energy states in Eq. (\ref{eq: Ising-Dicke implementation}) are the following~\cite{Steck-2023}: $\omega_1 = 6.835 \text{ GHz}$, $ \omega_d - \omega_e = 814.5 \text{ MHz}$. We consider pump frequencies detuned from the atomic transitions, with detunings much larger than the hyperfine splittings, namely, $\omega_{\Delta d}= \omega_d - \Delta_d$ and $\omega_{\Delta e}= \omega_e - \Delta_e$, with $\Delta_d = 100 \text{ GHz}$ and $\Delta_e = \Delta_d -(\omega_d - \omega_e)+ 2(\omega_1 - \Delta_1) = 113 \text{ GHz}$. From these, we can extract $\Delta_1$, which is related to $\omega_z$ via Eq. (\ref{omega_z correspondence 2}). The cavity is also detuned from the atomic transition as $\omega_c = \omega_d - 127\text{ GHz}$. 

Next, we estimate the effective light-atom couplings $\lambda_{d/e}$ in Eq. (\ref{eq: eff Hamiltonian}). The effective Dicke model requires a constraint on the Rabi frequencies. Therefore, we consider $\Omega_d = \Omega_e = 0.5 \text{ MHz}$. Assuming $N=2\times 10^5$ atoms in the cavity and an effective cavity frequency $\omega_a = 1.0 \text{ MHz}$, we use Eqs. (\ref{eq: lambdas and Veff}) and (\ref{omega_z correspondence 1}) to estimate $\kappa$ and $g_{e/d}$, as well as the light-atom couplings $\lambda_d = 158.8\text{ kHz}$ and $\lambda_e = 149.5\text{ kHz}$, which fall within experimental reach. Note that $g = 2\lambda_d \approx 2\lambda_e \approx 0.3 \text{ MHz}$. Finally, we note that in Ref. \cite{Zhiqiang:17}, an implementation of the spin-$1$ Dicke model found coupling strengths $g = (0.05 \text{-} 0.2) \text{ MHz}$, and a superradiant phase transition at approximately $g_c \approx 0.1 \text{ MHz}$ (see Fig. 2 in that reference).

We then turn to the Rydberg state and transitions. In Ref. \cite{Bernien:2017aa}, the authors simulate spin-spin interactions between Rydberg atoms using tweezer laser technology. They find a typical value $\Omega_\text{Ryd} = 2 \text{ MHz}$ and $\Delta_\text{Ryd} = (0\text{-}20) \text{ MHz}$, therefore $\Omega_{\text{Ryd}}^{2}/(4\Delta_{\text{Ryd}}) \geq 0.05 \text{ MHz}$. They also find coupling strengths between next-neighbors Rydberg atoms within the range $(24 \text{-} 1536) \text{ MHz}$, depending on the distance between atoms. Similar results for coupling strengths, Rabi frequencies, and detuning were reported in Refs. \cite{Zeiher-2016,PhysRevLett.128.113602}. For general purposes, in Table \ref{tbl: parameters}, we consider excitation to the Rydberg state $|70 S_{1/2}\rangle \equiv |\text{Ryd}\rangle$ via the intermediate state $|6^2 P_{3/2}, F=3\rangle$, as in Ref. \cite{Bernien:2017aa}.

With these estimations, we analyze the Ising-Dicke Hamiltonian in Eq. (\ref{app: QTFMI}). First, we note that the parameters in the main text are given in units of $J_z$, so the values in the phase diagram in Fig. \ref{fig: phase diagram QTFIM} are better compared to the literature when divided by the exchange coupling, i.e., $\omega_z/J_z$ and $(g/J_z)^2$. Thus, the phase transition presented in Fig. \ref{fig: phase diagram QTFIM} is not independent of $J_z$. For example, assuming $g_c = 0.1 \text{ MHz}$, the critical value $(g_c/J_z)^2 = 4$ corresponds to $J_z = 0.05 \text{ MHz}$, and a longitudinal field $\omega_z = -0.03 \text{ MHz}$, thus $\omega_z / J_z = -0.6$. To conclude, our analysis suggests that an effective coupling as weak as $50$ kHz is sufficient to achieve the results presented here. We summarize these results in Table \ref{tbl: parameters}.

\begin{table}
\hfill{}%
\begin{tabular}{|c|c||c|c|}
\hline 
Parameter & Value ($\times\, 2\pi$) & Parameter & Value ($\times\, 2\pi$) \tabularnewline
\hline 
\hline 
$\omega_{c}$ & $376.985$ THz & $\Delta_{c}$ & $127$ GHz\tabularnewline
\hline 
$\omega_{d}$ & $377.112$ THz & $\Delta_{d}$ & $100$ GHz\tabularnewline
\hline 
$\omega_{e}$ & $377.111$ THz & $\Delta_{e}$ & $112.854$ GHz\tabularnewline
\hline 
$\omega_{1}$ & $6.835$ GHz & $\Delta_{\text{Ryd}}$ & $20$ MHz\tabularnewline
\hline 
$\omega_{\Delta d}$ & $377.012$ THz & $\Delta_{1}$ & $0.22$ MHz\tabularnewline
\hline 
$\omega_{\Delta e}$ & $376.998$ THz & $\tilde{\omega}_{a}$ & $-20.165$ GHz\tabularnewline
\hline 
$\omega_{\Delta r}$ & $1009.014$ THz & $\omega_{1}'$ & $6.834$ GHz\tabularnewline
\hline 
$\Omega_{d}$ & $0.5$ MHz & $\lambda_{d}$ & $0.159$ MHz\tabularnewline
\hline 
$\Omega_{e}$ & $0.5$ MHz & $\lambda_{e}$ & $0.149$ MHz\tabularnewline
\hline 
$\Omega_{\text{Ryd}}$ & $2$ MHz & $g$ & $0.308$ MHz\tabularnewline
\hline 
$g_{d}$ & $142.009$ MHz & $\omega_{a}$ & $1.0$ MHz\tabularnewline
\hline 
$g_{e}$ & $150.860$ MHz & $J_{z}$ & $0.05$ MHz\tabularnewline
\hline 
$N$ & $2\times10^{5}$ & $\omega_{z}$ & $-0.03$ MHz\tabularnewline
\hline 
\end{tabular}\hfill{}

\caption{Parameters for the Hamiltonian of Eq. (\ref{eq: Ising-Dicke implementation}) and the $H_\text{QTFIM}$ of Eq. (\ref{app: QTFMI}).}

\label{tbl: parameters}
\end{table}

\end{document}